\RequirePackage{fix-cm}
\documentclass[smallextended]{svjour3}       
\smartqed  
\usepackage{graphicx}
\usepackage[utf8]{inputenc}
\usepackage{amsmath}
\usepackage{amssymb}
\usepackage{color}
\usepackage{comment}
\usepackage{url}
\usepackage{tikz}
\usetikzlibrary{patterns}
\usepackage{subfig}

\newcommand{\prob}[1]{{\sc{#1}}}

\begin{document}


\title{Enhancing Security via Deliberate Unpredictability of Solutions in Optimisation}


\author{Daniel Karapetyan \and Andrew J. Parkes}


\institute{D. Karapetyan \at
              School of Computer Science, University of Nottingham \\
              \email{daniel.karapetyan@nottingham.ac.uk}
           \and
            A. J. Parkes \at
              School of Computer Science, University of Nottingham \\
              \email{andrew.parkes@nottingham.ac.uk}}

\date{Received: 1 Mar 2020 / Accepted: 23 May 2020}

\maketitle

\begin{abstract}
    The main aim of decision support systems is to find solutions that satisfy user requirements.
    Often, this leads to predictability of those solutions, in the sense that having the input data and the model, an adversary or enemy can predict to a great extent the solution produced by your decision support system.
    Such predictability can be undesirable, for example, in military or security timetabling, or applications that require anonymity.
    In this paper, we discuss the notion of solution predictability and introduce potential mechanisms to intentionally avoid it.
    
\end{abstract}


\section{Introduction}

A search algorithm, even non-deterministic, is likely to be biased to some solutions, and hence anyone knowing the input data and the algorithm might be able to predict much of the solution.
This can be an issue if the solution has to be kept in secret.
One of many examples of this is the scheduling of tasks in cloud computing.
In this problem, computational tasks are assigned to various servers and time slots.
While respecting the constraints and efficiency considerations, one may want to keep the schedule as unpredictable as possible to reduce the chances of the potential intruder to guess the server and/or time slot assigned to a specific task.

Solution unpredictability can be understood in many ways.
In our example, one may be interested in predicting the exact server and time slot for a task, or may be interested in predicting only the server, or even an approximate location of the server (i.e., the specific data centre).
Our aim here is not to give a generic formulation of the problem; but to point out potential interesting extensions of the classic decision support, and to provide some relevant results, and so to encourage further discussion of the topic.

In particular, we discuss diversity issues in the context of the assignment problem, or specifically of variations of Perfect Matching Problems in bipartite graphs, which is closely related to the task assignment in cloud computing.
We want the locations/times of different tasks to be unpredictable (to make hacking harder) and so need diverse assignments to select from.

\section{Generation of Unpredictable Solutions}

\begin{figure}[bt]
    \centering
    \begin{tikzpicture}
    \draw[pattern=dots, fill opacity=0.2, text opacity=1] (-3, -2) rectangle (6, 2.0) node[midway, above=1.5cm] {Solution space};
    \draw[fill=red, fill opacity=0.2, text opacity=1] (0, 0) circle (1cm) node {$A$};
    \draw[fill=blue, fill opacity=0.2, text opacity=1] (4, 0) circle (0.4cm) node {$B$};
    \draw[thick, ->] (-3.5, -3) -- (6.5, -3) node[anchor=south east] {Decision variable};
    \end{tikzpicture}
    
    \caption{
        Example of solution space with all the feasible solutions grouped into two clusters $A$ and $B$.
        Cluster $A$ contains many more solutions than cluster $B$; as a result, uniform sampling from the entire set of feasible solutions will be biased towards cluster $A$, hence the value of the `decision variable' will be `predictable'.}
    \label{fig:clusters}
\end{figure}
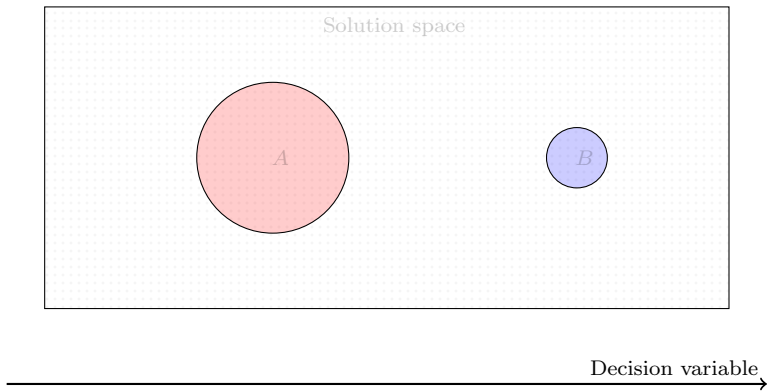

A straightforward approach to achieve unpredictability of solutions is to randomly sample the set of feasible solutions.
(A standard technique to do this is the ``rapidly mixing Markov Chains'', e.g. see \cite{guruswami2016rapidly,SinclairJerrum1989:rapid-mixing-MCs})
However, to enhance security, the sampling should not necessarily be uniform.
To illustrate this, we refer to Figure~\ref{fig:clusters}.
in which the sets of feasible solutions form clusters, i.e. subsets of similar solutions (e.g. see \cite{Parkes1997:clustering}).
In the example given in Figure~\ref{fig:clusters}, all the solutions are grouped into two clusters: $A$ and $B$.
Hence, solutions within each cluster share similar values of the `decision variable'.
Cluster $A$ contains more solutions, and hence a solution selected with uniform sampling is likely to be from cluster $A$\@.  
This will make the value of the decision variable predictable; with high probability, it will correspond to the first cluster.

To address this issue, we may want to pre-select a subset of diverse feasible solutions and then pick one of them randomly.
For example, such a subset may be obtained by selecting 10 feasible solutions such that the total Hamming distance between them is maximised.
(In a loose sense, we are doing the exact opposite of the work on minimal perturbations, such as \cite{MuellerEtall2005:PATAT:minimal}, which aimed to find nearby solutions, and instead are looking for ``maximal perturbations''.)

This approach is likely to generate interesting optimisation challenges.
Doing it directly by first enumerating all solutions is generally impractical, even for problems where finding a solution is easy. 
Indeed, just counting all the feasible solutions is generally \#P-hard, and the solutions sets are typically exponential in size.

Also, there is the challenge of selecting diverse sets of solutions.
This corresponds to a kind of ``Maximum Diversity Problem'' and again likely to be NP-hard.
We expect that heuristic approaches can be used to address these complexity issues, though maybe with special cases for which efficient algorithms are available.

However, in this work-in-progress paper, naturally, we do not answer these questions.
Instead we consider the relatively simple case of perfect matching problems, or assignment problems; though these problems are of interest in their own right. 
For enhanced security of systems using assignment problems, we might well want to increase the unpredictability of such solutions. 
Accordingly, in the next section, Section~\ref{sec:perfect-matching}, 
we study the problem of finding diverse solutions to the perfect matching problem, and present relevant decision and optimisation problems, with some initial work on solution methods.

\section{Diverse Solutions Sets for the Perfect Matching Problem}
\label{sec:perfect-matching}

Firstly, a quick reminder of the base problem:

\begin{description}
\item[NAME:] \prob{Perfect-Matching}

\item[INSTANCE:] A bipartite graph $G = (U,V,E)$ over vertices $(U,V)$ of sizes $(n,n)$ and with edges $E$.

\item[SOLUTION:] A vertex-disjoint subset $M \subseteq E$ of $n$ edges, or equivalently, a subset of edges such that it covers every node and the edges in this subset are disjoint (do not share any nodes).
\end{description}

Finding one solution (perfect matching) is well-known to be polynomial-time (e.g. using the Hungarian method). 
However, this does not mean that all questions about perfect matchings are necessarily easy. 
For example, counting the number of solutions is \#P (sharp-P); a class that is (generally assumed) much harder than NP. 
In particular, we remark, that there may well be that there are questions, relevant to diversity, about the set of solutions (perfect matchings) that may also be harder than P (under usual assumptions, such as P $\neq$ NP).

A problem that naturally occurs when studying diverse solutions to the matching problem is to find a pair of well-separated perfect matchings.
We will define separation as the number of edges that are not shared by the two matchings. 
The corresponding decision problem is as follows.

\begin{description}
\item[DEFINITION:] \prob{Maximum-Separated-Perfect-Matchings}

\item[INSTANCE:] ~
\begin{itemize}
\item Bipartite (unweighted) graph $G$ on $(n,n)$ nodes;
\item Integer $0 \le d \le n$.
\end{itemize}

\item[QUESTION:]
Do there exist perfect matchings $M_1$ and $M_2$, such that $M_1$ and $M_2$ differ on at least $d$ assignments?
\end{description}

We do not know the complexity of this problem and leave it as an open question.
Below, however, we show that several related problems are polytime.

Firstly, suppose that we are given one perfect matching, and to promote diversity, we want to find another one that is ``as different as possible''.
The corresponding decision problem is as follows.

\begin{description}
\item[DEFINITION:] \prob{Distant-Perfect-Matching}

\item[INSTANCE:] ~
\begin{itemize}
\item Bipartite (unweighted) graph $G$, on $(n,n)$ nodes;
\item Perfect matching $M_1$;
\item Integer $0 \le d \le n$.
\end{itemize}

\item[QUESTION:]
Does there exist another perfect matching $M_2$, such that $M_1$ and $M_2$ differ on at least $d$ assignments?
In other words, does there exist a matching $M_2$ such that $|M_1 \cap M_2| \le n - d$?
\end{description}

The maximum distance, $d = n$, is easy because it means that no edges can be shared. 
Hence, we can simply solve this case by removing all the edges in $M_1$ from $E$, and then looking for a perfect matching in this reduced graph.

The problem of finding maximum $d$ can also be solved in poly-time using the given solution $M_1$ to modify the weights of the edges, giving a new weighted graph and then doing a maximum weight perfect matching on this graph.

This approach has the drawback that we need to provide the first matching.
Instead, generally we want to simultaneously find a pair of well-separated perfect matchings -- ones differing on at least $d$ edges.
Using the usual distinction between ``maximal'' (local) and ``maximum'' (global), this leads to two problems, Firstly, the ``maxim\textbf{al}'' separation:

\begin{description}
\item[DEFINITION:] \prob{Maximal-Separated-Perfect-Matchings}

\item[INSTANCE:] ~
\begin{itemize}
\item Bipartite (unweighted) graph $G$ on $(n,n)$ nodes;
\end{itemize}

\item[TASK:]
Find a pair of matchings $M_1$ and $M_2$ such that no matching is further from $M_1$ than $M_2$ is, and vice versa.
\end{description}

This is in poly-time because we just iterate solution to \prob{Distant-Perfect-Matching}, switching between which matching is considered the fixed one. 
Starting from any matching, call it $M_1$, then find the most distant, call it $M_2$, then find the most distant from $M_2$ etc., terminating when the distance no longer increases -- which must happen within $O(n)$ iterations.

Another special case of the \prob{Maximum-Separated-Perfect-Matchings} is when $d = n$, which requires a disjoint pair of perfect matchings. 
Consider a subgraph of $G$ with the edge set $M_1 \cup M_2$.
Each vertex now has two distinct edges to it; by following these we get disjoint cycles. 
So the $d=n$ case is equivalent to finding a ``Disjoint Vertex Cycle Cover'' -- a set of disjoint cycles that contain all the vertices. 
This is known to be polynomial time by conversion to a matching problem\footnote{See \url{https://en.wikipedia.org/wiki/Vertex_cycle_cover}}.

\section{Conclusions}
\label{sec:conclusions}

In this paper, we discussed the concept of unpredictability of solutions in automated decision support.
As a motivating example, we consider a simple assignment problem, which could easily be part of task scheduling in cloud computing.
The aim is that unpredictability of task assignments will increase security of the system, by making it harder for malicious agents to guess locations and time slots of tasks.
The most obvious approach to achieving unpredictability, random sampling of solutions, turns out to be computationally hard and weak.
Indeed, uniform sampling is both complex and would not necessarily give us the desired diversity of solutions.

Hence, we focus on finding a few diverse solutions; then we can select one of them randomly to achieve unpredictability.
We model the task scheduling using the bipartite matching.
We gave some initial definitions that relate to finding diverse pairs of matchings. 
In particular, observing that finding maximally separated matchings is possible in polynomial time.
Though not (yet) answering the question of finding the pairs with maximum separation.
Also whilst we found that it is easy to find a \textbf{pair} of non-overlapping solutions, we do not know whether this result generalises to a higher number of solutions.

This is still work in progress and more research is needed to establish efficient methods for increasing unpredictability of solutions in new and existing decision support systems.  
For example, here we have discussed only issues of selecting from `feasible', but it could be that this includes quality being above some threshold.

\bibliographystyle{spmpsci}
\bibliography{refs}

\begin{thebibliography}{1}
\providecommand{\url}[1]{{#1}}
\providecommand{\urlprefix}{URL }
\expandafter\ifx\csname urlstyle\endcsname\relax
  \providecommand{\doi}[1]{DOI~\discretionary{}{}{}#1}\else
  \providecommand{\doi}{DOI~\discretionary{}{}{}\begingroup
  \urlstyle{rm}\Url}\fi

\bibitem{guruswami2016rapidly}
Guruswami, V.: Rapidly mixing markov chains: A comparison of techniques (a
  survey) (2016)

\bibitem{MuellerEtall2005:PATAT:minimal}
M{\"u}ller, T., Rudov{\'a}, H., Bart{\'a}k, R.: Minimal perturbation problem in
  course timetabling.
\newblock In: E.~Burke, M.~Trick (eds.) Practice and Theory of Automated
  Timetabling V, pp. 126--146. Springer Berlin Heidelberg, Berlin, Heidelberg
  (2005)

\bibitem{Parkes1997:clustering}
Parkes, A.J.: Clustering at the phase transition.
\newblock In: Proceedings of the fourteenth national conference on artificial
  intelligence and ninth conference on Innovative applications of artificial
  intelligence (AAAI-97), pp. 340--345 (1997)

\bibitem{SinclairJerrum1989:rapid-mixing-MCs}
Sinclair, A., Jerrum, M.: Approximate counting, uniform generation and rapidly
  mixing markov chains.
\newblock Information and Computation \textbf{82}(1), 93 -- 133 (1989).
\newblock \doi{https://doi.org/10.1016/0890-5401(89)90067-9}

\end{thebibliography}

\end{document}